\newcommand{\beq}{\begin{equation}}
\newcommand{\eeq}{\end{equation}}

\documentclass{revtex4}
 
\begin{document}

\title{Grand Unification with and without Supersymmetry  \footnote{Talk given at the VI SILAFAE, Puerto Vallarta, M\'exico, November 2006}}
 \author{Alejandra Melfo}

\address{CFF, Universidad de Los Andes, M\'erida, Venezuela, and Institute J. Stefan, Ljubljana, Slovenia.}

\begin{abstract}

Grand Unified Theories based on the group SO(10) generically provide interesting and testable relations between the charged fermions and neutrino sector masses and mixings. In the light of the recent neutrino data, we reexamine these relations both in supersymmetric and non-supersymmetric models, and give a brief review of their present status.
 
 \end{abstract}

\maketitle
 
\section{Introduction}

Of the two main candidates for a unifying gauge group, $SU(5)$ and $SO(10)$, the latter is often cited as the one most suitable in accounting for neutrino masses. This is not at all clear, in particular after the recent suggestion of a non-supersymmetric $SU(5)$ model with a Type III see-saw mechanism, that provides not only a consistent unified theory with neutrino masses, but also predictions directly testable at the LHC \cite{bajcsenj}.  $SU(5)$ models are  in addition generally simpler, making proton decay rate  calculations  feasible. But the remarkable feature of  $SO(10)$  is that   the 16-dimensional spinor representation can accommodate a complete family of fermions, including the right-handed neutrino.  This complete unification of quarks and leptons opens up the possibility of obtaining connections between the charged fermions and the neutrino sector. Furthermore, $SO(10)$ has the Left-Right group as a subgroup \cite{leftright}, making the implementation of the see-saw mechanism \cite{seesaw} very natural in these theories. 
 
The smallness of neutrino mass is also calling for an intermediate scale. Namely, the see-saw mechanism will give neutrinos  a small mass after the electroweak symmetry breaking, provided the $B-L$ symmetry is broken at some scale $M_R$, of order $M_W^2/M_R$. With the current limit on $\Delta m_{23}^2 \simeq 2.5 \times 10^{-3} eV^2$, the mass of the heaviest neutrino  is an indication that $M_R \sim 10^{13} GeV$, below $M_{GUT} \sim 10^{16} GeV$, and even smaller in the case neutrinos are degenerate. It is only an indication, of course, since three orders of magnitude can in principle be accounted for by  the Yukawa coupling constants of the right-handed neutrinos.  In practice however it is not so simple, as has been recently seen in the case of the Minimal Supersymmetric Grand Unified Model, which fails to predict a large enough mass for the heaviest neutrino, as we will review below. 

In their turn, intermediate scales are an indication of the possibility of a non-supersymmetric unification. The Minimal Supersymmetric Standard Model (MSSM) gives one-step unification at $M_{GUT}$ with such a good precision \cite{unification} that it leaves basically no room for intermediate scales, if the scale of supersymmetry breaking is reasonably close to $M_W$, as required by the supersymmetric solution to the hierarchy problem. Assuming an intermediate scale of symmetry breaking allows for unification without supersymmetry, and it is possible in many cases to have this intermediate scale at the required value, as we will see below.  

In this talk, we review briefly the current status of supersymmetric and non-supersymmetric theories based on $SO(10)$, with emphasis on the Yukawa sector and the possibility of general relations connecting the charged fermion and neutrino sectors.

\section{Yukawa sector}

In $SO(10)$, the Yukawa sector is particularly simple. Each family of fermions is in a {\bf 16}-dimensional spinor representation, and they can therefore couple to only three fields,
  $\bf{10_H}$, $\bf{120_H}$ and ${\bf\overline{126}_H}$ 
representations, since 

\beq 
{\bf 16\times 16=10+120+126}\;.
\eeq
 Furthermore, the Yukawa coupling with $120_H$ is antisymmetric in family space.  The Higgs fields decompose under the SU(2)$_L\times$ SU(2)$_R\times$ SU(4)$_C$ Pati-Salam \cite{Pati:1974yy} group as
\begin{eqnarray}
{\bf 10} &=& (2,2,1) + (1,1,6)  \nonumber \\
{\bf\overline{126}} &=& (1,3,10) + (3,1,\overline{10}) + (2,2,15) + (1,1,6) \\
{\bf 120} &=& (1,3,6) + (3,1,6) + (2,2,15) + (2,2,1) + (1,1,20)\nonumber
\end{eqnarray}

The ${\bf 126_H}$ provides mass terms for 
right-handed and left-handed neutrinos:

\begin{equation}
M_{\nu_R}=  \langle 1,3,10 \rangle  \,Y_{126},\ \ 
M_{\nu_L}=  \langle 3,1,\overline{10}\rangle  \, Y_{126} 
\label{mlr}
\end{equation}

\noindent
which means that one has both type I and type II seesaw:

\beq 
M_N= -M_{\nu_D}M_{\nu_R}^{-1}M_{\nu_D}+M_{\nu_L}\;
\label{mnu}
\eeq

\noindent
In the type I case it is the large vev of $(1,3,10)$ that provides the 
masses of right-handed neutrino whereas in the type II case, the 
left-handed triplet provides directly light neutrino masses through a 
small vev \cite{Lazarides:1980nt,Mohapatra:1980yp}. 
The disentangling of the two contributions is in general hard. 

One can now ask: what is the minimal Yukawa sector one can choose? Clearly only one of the  three representations cannot do the job, as then one would have a fixed relation between the down quarks and the charged leptons for all generations.  

If one is interested in a   renormalizable version of the see-saw 
mechanism the representation 
${\bf \overline{126}_H}$ is indispensable, since it breaks the SU(2)$_R$ 
group and gives a see-saw neutrino mass.   
By itself it gives no fermionic 
mixing, so it does not suffice. The realistic fermionic spectrum 
requires adding either $\bf{10_H}$ or $\bf{120_H}$.  An interesting possibility which we shall not review here is a radiatively-induced see-saw 
 \cite{Bajc:2005aq}. 
 
  Let us start by analyzing the non-supersymmetric case, with an extra ${\bf 10_H}$ field \cite{Bajc:2005zf}.  The most general Yukawa interaction is  
\begin{equation}
\label{yukawa}
{\cal L}_Y={\bf 16_F}\left({\bf 10_H} Y_{10}+
{\bf \overline{126}_H} Y_{126}\right){\bf
16_F}+h.c.\;.
\end{equation}
\noindent
where $Y_{10}$ and $Y_{126}$ are symmetric matrices in the generation space. 
With this one obtains relations for the Dirac fermion masses 
\begin{eqnarray}
M_D=M_1+M_0&,\;&M_U=c_1M_1+c_0 M_0\;\;,\nonumber\\
M_E=-3M_1+  M_0& , \;&M_{\nu_D}=-3c_1M_1+ c_0 M_0,
\label{mod1}
\end{eqnarray}
\noindent
where we have defined
\begin{equation}
M_1= \langle 2,2,15 \rangle_{126}^d  \, Y_{126}\;, \quad
M_0= \langle 2,2,1 \rangle_{10}^d  \, Y_{10} \; ,
\label{m1m0}
\end{equation}
\noindent
and
\begin{equation}
c_0=\frac{\langle 2,2,1\rangle_{10}^u }{ 
\langle 2,2,1\rangle_{10}^d } \, , \;
 c_1=\frac{\langle 2,2,15\rangle_{126}^u }{
\langle 2,2,15\rangle_{126}^d }\;.
\label{c1c0}
\end{equation}

\noindent
 In the physically sensible approximation 
$\theta_q= V_{cb}=0$, these relations imply

\begin{equation}
c_0=\frac{m_c(m_\tau-m_b)-m_t(m_\mu-m_s)}{m_sm_\tau-m_\mu m_b}
\approx \frac{m_t}{m_b}\;,
\end{equation}

Notice that this means that  ${\bf 10_H}$ cannot be real, since in that case one would have 
$|\langle 2,2,1\rangle_{10}^u|= 
|\langle2,2,1\rangle_{10}^d|$, implying $m_t/m_b$ of order one. It is necessary to complexify ${\bf 10_H}$, just as in a supersymmetric theory. If taking advantage of this fact one decides to impose a Peccei-Quinn symmetry, thus providing a Dark Matter candidate, the Yukawa sector in non-supersymmetric and supersymmetric models is similar. 

In this case, this model has the interesting feature of automatic connection between $b-\tau$ unification 
and large atmospheric mixing angle in the type II see-saw. 
From $M_{\nu_L}\propto Y_{126}\;,$ one has $M_{\nu_L}\propto M_D-M_E$.
as shown in \cite{Bajc:2002iw,Bajc:2004fj}.  This fact has inspired the careful 
study of the analogous supersymmetric version where $m_\tau \simeq m_b$ 
at the GUT scale works rather well \cite{Babu:2005ia} . In the non-supersymmetric theory, 
$b-\tau$ unification no longer happens,  one has instead $m_\tau \sim  2 m_b$ \cite{Arason:1992eb}. 
The realistic theory will require a Type I seesaw, or an admixture of 
both possibilities.

Suppose now that  we choose instead $\bf{120_H}$  \cite{Bajc:2005zf}. Since $Y_{120}$ 
is antisymmetric, this means only 3 new complex couplings on top of 
$Y_{126}$. One gets in this case
\begin{eqnarray}
M_D=M_1+M_2&\;\;,\;\;&M_U=c_1M_1+c_2M_2\;\;, \\
M_E=-3M_1+c_3M_2&\;\;,\;\;&M_{\nu_D}=-3c_1M_1+c_4M_2\; \nonumber
\end{eqnarray}
\noindent
where $M_1$ and $c_1$ are defined in (\ref{m1m0}),(\ref{c1c0}), and:
\begin{eqnarray}
M_2=Y_{120}\left(\langle 2,2,1 \rangle_{120}^d\ +
\langle 2,2,15 \rangle_{120}^d \right)\; & ,&
c_2 = \frac{\langle 2,2,1 \rangle_{120}^u +
\langle 2,2,15 \rangle_{120}^u }{\langle 2,2,1 \rangle_{120}^d +
\langle 2,2,15 \rangle_{120}^d }\;, \nonumber  \\
c_3={\langle 2,2,1\rangle_{120}^d-3
\langle 2,2,15\rangle_{120}^d\over 
\langle 2,2,1\rangle_{120}^d+
\langle 2,2,15\rangle_{120}^d}\; &, &
c_4={\langle 2,2,1\rangle_{120}^u-3
\langle 2,2,15\rangle_{120}^u\over 
\langle 2,2,1\rangle_{120}^d+
\langle 2,2,15\rangle_{120}^d}\;.
\label{lec}
\end{eqnarray}
It is easy to see that again there is a need to complexify the Higgs fields, by arguments similar to  the case of ${\bf 10_H}$. 

 In order to obtain algebraic expressions, from which a clearer physical meaning can be extracted, one can restrict the analysis   to  the second and third 
generations. Later, numerical studies could include the effects of the first generation as a perturbation.  In the basis where $M_1$ is diagonal, real and non-negative, for the two-generation case one gets:
\begin{equation}
M_1\propto 
\left(
\begin{array}{cc}
\sin^2\theta & 0 \\
0 & \cos^2\theta
\label{m1}
\end{array}
\right)
\label{deag}
\end{equation}
\noindent
and the most general charged fermion matrix 
can be written as:
\begin{equation}
M_f=\mu_f 
\left(
\begin{array}{cc}
\sin^2\theta & i (\sin\theta \cos\theta+\epsilon_f) \\
-i (\sin\theta \cos\theta+\epsilon_f) & \cos^2\theta
\end{array}
\right)\;,
\label{eqq2}
\end{equation}

\noindent
where $f=D,U,E$ stands for charged fermions. $\epsilon_f$ is a small ratio that vanishes 
for negligible second generation masses, {\it i.e.} 
$|\epsilon_f|\propto m_2^f/m_3^f$. The real parameter $\mu_f$ sets the third generation mass scale.
By calculating up to leading order in $|\epsilon_f|$, we have to the following interesting predictions  \cite{Bajc:2005zf}:

\begin{enumerate}

\item  type I 
and type II seesaw lead to the same structure  

\begin{equation}
M_N^I\propto M_N^{II}\propto M_1 
\end{equation}

\noindent
so that in this basis the neutrino mass matrix is diagonal.
The angle $\theta$ is identified with the leptonic 
(atmospheric) mixing angle $\theta_A$ up to terms of the order 
of $|\epsilon_E|\approx m_\mu/m_\tau$. 
For the neutrino masses we obtain from (\ref{m1})

\begin{equation}
\label{dmsm}
\frac{m_3^2-m_2^2}{m_3^2+m_2^2}=
\frac{\cos2\theta_A}{1-\sin^22\theta_A/2} + {\cal O}(|\epsilon|)
\end{equation}  

\noindent
which provides an interesting  connection between 
 the degeneracy 
of neutrino masses and  the maximality 
of the atmospheric mixing angle.

\item  the ratio of tau and bottom mass at 
the GUT scale is given by: 

\begin{equation}
\frac{m_\tau}{m_b}=3 + 3 \sin 2\theta_A\ {\rm Re}[\epsilon_E-\epsilon_D]
+{\cal O}(|\epsilon^2|)
\end{equation}

\noindent

This is not correct in principle, the extrapolation in 
  standard model gives
$m_\tau\approx 2 m_b$. However,  several effects modify this 
conclusion, such as for example the inclusion of the first generation or the running of Yukawa couplings. We would in any case expect that $m_b$ comes out 
as small as possible.

\item   the 
quark mixing is found to be:

\begin{equation}
|V_{cb}|= |\ {\rm Re}\xi -
i\cos2\theta_A\ {\rm Im}\xi  |+ {\cal O}(|\epsilon^2|)
\end{equation} 

\noindent
where $\xi=\cos2\theta_A\  (\epsilon_D-\epsilon_U)$.
This equation demonstrates the successful coexistence of small and 
large mixing angles. In order for it to work quantitatively, 
$|\cos{2\theta_A}|$ should be as large as possible, i.e.~$\theta_A$ 
should be as far as possible from the maximal value $45^\circ$. 
 To make a definite numerical statement, again, the  effects 
from the first generation and the loops  have to be included.

\end{enumerate}

\section{Unification constraints}

In the non-supersymmetric case,  an intermediate scale is necessary for  unification to succeed, and in
the over-constrained models discussed here, the Dirac neutrino 
Yukawa couplings are not arbitrary. Thus one must make sure that the 
pattern of intermediate mass scale is consistent with a see-saw mechanism
for neutrino masses. The lower limit on $M_R$ stems from the heaviest neutrino mass
$ m_\nu \geq m_t^2/M_R $,
which gives $M_R \geq 10^{13}$ GeV or so. One can now turn 
to the useful table of Ref.~\cite{Deshpande:1992em}, where the most 
general patterns of SO(10) symmetry breaking with two intermediate 
scales consistent with proton decay limits are presented.  Most models are ruled out by the constrain above; the most promising candidates are those 
with an intermediate SU(2)$_L\times $SU(2)$_R\times$ SU(4)$_C\times $P  
symmetry breaking scale (that is, the Pati-Salam group with unbroken 
parity). 

The supersymmetric models have been extensively studied in the last few years, in particular in the minimal version the so-called Minimal Supersymmetric GUT (MSGUT). Besides the ${\bf 10_H}$ and ${\bf \overline{126}_H}$ providing the fermion masses, in this model just one additional representation suffices to break the symmetry all the way to the Left-Right group (recall that ${\bf \overline{126}_H}$ is in charge of breaking it further to the Standard Model group) \cite{oldso10,plb}. It has been studied in detail, and the mass spectrum calculated simultaneously by various groups \cite{aarti,fuk,melfo}. It has a number of very interesting features, like the conservation of R-parity at all energies, providing the LSP as candidate for dark matter \cite{rparity}, and the connection between the $b-\tau$ unification and a large atmospheric mixing angle \cite{Bajc:2002iw}.

The MSGUT has so few parameters, 26 in all, that is becomes even too predictive for its own good. Namely, the composition of the light Higgs doublets is no longer arbitrary, but given in terms of very few parameters. In fact, the dependence is such that only one complex parameter is truly relevant for the fitting of the known fermion masses and mixing, in particular for neutrino masses. 

The superpotential of the model  is 
(see e.g., \cite{plb} for a more explicit notation): 
\begin{eqnarray}
W &=& {\bf 16_m ( {\it Y_{\bf 10}}\ 10_H +  {\it Y_{\bf\overline{126}}}\ 
\overline{126}_H) 16_m}   + m {\bf 210_H}^2 + \lambda {\bf 210_H}^3
+ M {\bf 126_H\ \overline{126}_H} \nonumber \\
& &+ \eta  {\bf 126_H\ 210_H\ \overline{126}_H} + 
m_H {\bf 10_H}^2 + {\bf 210_H\ 10_H}\ 
(\alpha {\bf 126_H}+\overline\alpha {\bf \overline{126}_H})
\label{supot}
\end{eqnarray}
 where 
  $Y_{\bf 10}$ and $Y_{\bf\overline{126}}$ are 
the two Yukawa couplings of the theory. 
The rest of the superpotential describes 
the Higgs sector at high (GUT) scale.
  
 Notice that in fact, this model has only one mass scale. 
The mass $m_H$ is fixed by the fine-tuning \cite{plb,melfo} condition:
\begin{equation}
m_H=m\times \frac{\alpha\overline\alpha}{2 \eta \lambda}\ 
f(x)
\end{equation}
with $f(x)$ a known function, that is needed to ensure   a
pair of light Higgs doublets $H_u$ and $H_d$
in the spectrum of the low energy supersymmetric standard model (MSSM).
The parameter $x$ measures one of the vacuum expectation values (VEV)  
in unities of $m/\lambda$ \cite{plb}. The mass $M$ can be calculated in terms of $m$ and the other 
parameters due to the consistency condition for symmetry breaking:

\begin{equation}
M=m\times \frac{\eta}{\lambda}\ \frac{3-14 x + 15 x^2-8x^3}{(x-1)^2}
\end{equation}
 which amounts  to a 
replacement of  $M$ in favor of $x$.
Thus, any new scale, as the right-hand\-ed neu\-tri\-no mass sca\-le, 
has to arise by the mass $m$ times
some adimensional quantity. One can 
then find the symmetry-breaking VEVs as functions of $x$ \cite{plb}. The SM singlets that acquire vevs are five:
$\langle 1,1,1 \rangle_{210}  , \langle 1,1,15 \rangle_{210} , \langle 1,3,15 \rangle_{210} , \langle 1,3,10 \rangle_{\overline{126}} $ and $ \langle 1,3, \overline{10} \rangle_{126}$ .
The Higgs sector has just 8 real parameters after fine-tuning, which can be chosen as e.g.:

\begin{equation}
m,\ \alpha,\ \overline{\alpha},\ |\lambda|,\ |\eta|,\
\phi=\mbox{arg}(\lambda)=-\mbox{arg}(\eta),\  
x=\mbox{Re}(x)+ i \mbox{Im}(x).
\end{equation}

The parameter $x$ is known to be convenient 
to describe the VEVs and the masses of the particles; 
the dependences of observed fermion masses 
on the other parameters is rather simple, 
whereas the behavior in $x$ is usually non trivial \cite{melfo}. Using this, it is found that 
because of the absence of an intermediate scale  $M_R$, the type II see-saw is readily seen to fail the test \cite{Bajc:2005qe,Aulakh:2005mw}, the neutrino masses come out just too small. Surprisingly enough, the same happens in the Type I case \cite{Aulakh:2005mw, Bertolini:2006pe}. Namely, if ${\bf 126_H}$ is to play any role in the charged fermion masses, its Yukawa couplings cannot be arbitrarily small. The fitting requires them to be quite large, and since the Type I see-saw mass is proportional to the inverse Yukawa matrix, again it turns out too small. The current status is that this model is not viable in its minimal form.

\section{The non-minimal models}

One obvious alternative for the supersymmetric case is to enlarge the Yukawa sector by adding a ${\bf 120_H}$, on top of ${\bf 10_H}$ and ${\bf \overline{126}_H}$. This means three  complex additional parameters form the (asymmetric) Yukawa matrix, along with five new complex parameters in the superpotential, one of which could be removed by a redefinition of the new field. Since the 120-dimensional representation contains no Standard Model singlets, the symmetry breaking pattern of the minimal model is maintained, which somehow simplifies the analysis. However, the parameter space is now so large that additional assumptions (such as the possibility of spontaneous CP violation) have to be made in order to attempt the fitting of fermion data.

Once the Yukawa sector becomes maximal, there is very little difference between the minimal model and the most simple alternative, the model with   54 and 45-dimensional representations instead of 210, thoroughly studied in \cite{nucphys}.  $ {\bf 54_H}$ and  ${\bf 45_H}$ fail to provide a coupling between ${\bf 10_H}$ and ${\bf \overline{126}_H}$, thus in order to have a rich enough Yukawa structure the simplest possibility is to add a ${\bf 120_H}$, in which case all the bidoublets mix. This version has only 5 parameters more than the minimal one. A detailed study of the supersymmetry breaking and particle spectrum has been carried out \cite{inprep}.  The superpotential has 5 more parameters than the extended MSGUT,  and the symmetry breaking conditions are given in terms of two mass ratios now.

It is reasonable to suppose that the maximal Yukawa sector can provide a good fitting of charged fermions and neutrinos. But this model has also the potential of providing a Type II dominance simply by fine-tuning a small mass for the left-handed triplet (a couple of orders of magnitude below $M_{GUT}$ would suffice). Unification constrains can still be satisfied, provided the triplet's contributions to the RGE is cancelled by other fields in the spectrum, with masses fine-tuned to approximately the same value.  
At one loop, the contribution to the RGE equations of the $SU(2)_L$ triplets $(1,3,\pm 1)$, can be cancelled out by two color sextets $(6,1,\pm 1/3)$ present in $126_H$, $\overline{126}_H$ and $120_H$, and a couple of doublets $(1,2,\pm 1/2)$, with the same quantum  numbers as the light MSSM Higgs. Thus, for example, the mass parameter of the ${\bf 10_H}$ can be fixed as usual, in order to fine-tune doublets at the electroweak scale. Then the mass of $\bf{126_H, \overline{126}_H}$, that of ${\bf 54_H}$ and that of ${\bf 120_H}$ can be fixed so as to have a light triplet, thus a type-II dominance in the neutrino mass. This means that the right-handed scale entering the neutrino mass becomes arbitrary again, depriving the GUT of one of its most important roles.

\section{Concluding remarks}

We have argued that the grand unified model based on the group $SO(10)$ provides, through quark-lepton unification, interesting connections between the charged and non charged fermion sectors. The see-saw mechanism, in its renormalizable version, calls for a ${\bf \overline{126}}$ representation, but it alone cannot accommodate the complete spectrum. Including a ${\bf 10_H}$ or ${\bf 120_H}$  cures this problem, but a complex field has to be used, or one is lead to wrong relations between quark masses  at the GUT scale. Complexifying the Yukawa sector, in turn, leads to the interesting possibility of invoking a Peccei-Quinn symmetry, thus allowing for the axion as a candidate dark matter. If the additional field is the ${\bf 10_H}$, one is lead to the connection between the large atmospheric angle and $b-\tau$ unification.  If on the other hand the antisymmetric ${\bf 120_H}$ is added, the two generation analysis suggests that a large atmospheric angle is related to small mixing between second and third generations of quarks, and that it prefers degenerate neutrinos.

Non-supersymmetric models, with an intermediate scale of at least $10^{13}$ GeV, satisfy the unification constrains in the case where the symmetry breaking proceeds through the Pati-Salam subgroup. Supersymmetric models do not allow for an intermediate scale, and that leads to too small neutrino masses in the case of Type II neutrino,  but even Type I neutrino masses come too small in the MSGUT, where the symmetry breaking requires only one additional ${\bf 210_H}$, and the Yukawa sector contains the minimal combination, ${\bf 10_H}$ and ${\bf \overline{126}_H}$. The restrictions coming form symmetry breaking and unification constraints prove fatal to the fitting of fermion masses. One is forced away from the minimal models into an enlarged  Yukawa sector, and then the predictivity of the models becomes  questionable. It is no longer clear if the extended MSGUT is to be preferred over the alternative model (with ${\bf 54_H}$ and ${\bf 45}$ ).

\section*{Acknowledgments}
  
  I wish to thank my collaborators Borut Bajc, Goran Senjanovi\'c, Francesco Vissani and Alba Ram\'irez. Thanks also to the organizers of the VI Silafae for a truly enjoyable meeting.

\end{document}